\documentclass[aps,pra,twocolumn,10pt]{revtex4-1}

\usepackage{natbib}
\usepackage{amsmath,amssymb,bm}
\usepackage{dsfont}
\usepackage{graphics}

\newcommand{\im}[0]{{\rm i}}

\newcommand{\ket}[1]{\left|{#1}\right\rangle}
\newcommand{\bra}[1]{\left\langle{#1}\right|}
\newcommand{\braket}[2]{\left\langle{#1}|{#2}\right\rangle}

\newcommand{\dbar}{d\hspace*{-0.08em}\bar{}\hspace*{0.1em}}

\begin{document}

\title{Combining spatio-temporal and particle-number degrees of freedom}

\author{Filippus S. \surname{Roux}}
\email{froux@nmisa.org}
\affiliation{National Metrology Institute of South Africa, Meiring Naud{\'e} Road, Brummeria, Pretoria, South Africa}
\affiliation{School of Physics, University of the Witwatersrand, Johannesburg 2000, South Africa}

\begin{abstract}
Quadrature bases that incorporate spatio-temporal degrees of freedom are derived as eigenstates of momentum dependent quadrature operators. The resulting bases are shown to be orthogonal for both the particle-number and spatio-temporal degrees of freedom. Using functional integration, we also demonstrate the completeness of these quadrature bases.
\end{abstract}

\maketitle

\section{Introduction}

Quantum information promises various new technologies, including quantum imaging \cite{ghost1,ghostdebate,ghostph}, quantum metrology \cite{maccone,escher,homsync2} and quantum communication \cite{gisin,bp2000,scarani}. Many of the systems that implement these quantum information technologies use quantum optics \cite{photonic}. In such cases, we often find that the physical setup predominantly employs either the spatial degrees of freedom \cite{zeiloam,malik,oamhom} or the particle-number degrees of freedom \cite{weedbrook,sqlight,cvteleport}. Where the spatial degrees of freedom are used, the implementation is often made in terms of spatial modes, such as orbital angular momentum modes \cite{allen,mair}. Those implementations that employ the particle-number degrees of freedom often use so-called continuous variables \cite{contvar1,contvar2}.

However, one cannot completely exclude the influence of other degrees of freedom from any quantum optics implementation. When spatial modes are use, one usually assumes that the states are single-photon states or two-photon states \cite{Birulaphot,sipephot,sr2}. Nevertheless, the laser sources and nonlinear crystals actually produce coherent states and squeezed states, thus polluting these low-particle-number states with higher particle-number states \cite{dusek,bunchou}. The effects of these undesirable contributions are treated as `noise' in the system.

On the other hand, continuous-variable systems always incorporate some spatio-temporal description of the optical field. In these implementations, multiple modes are allowed, but it is often assumed that these are fixed discrete modes \cite{contvar1,contvar2,brambilla,sqlight}. (Obviously, there are many exceptions, for instance, in the context of quantum imaging \cite{kolobov,ghostgauss}.) However, in practical systems, distortions and decoherence can introduce unwanted variability in these modes \cite{paterson,sr,qturb4} that may not be represented by the finite set of fixed modes.

A comprehensive approach that combines all the degrees of freedom allows one to perform thorough analyses of all such scenarios. It is especially relevant in cases where all these degrees of freedom are affected. Such a comprehensive approach requires the incorporation of both spatio-temporal and particle-number degrees of freedom into the analytic tools. If the current state of the art for implementations that focus on only one degree of freedom is able to achieve remarkable successes as they do, just imagine how powerful implementations that incorporate all these degrees of freedom would be \cite{sqltreps,lvovsky0,boydsupres,chille,supres,qsense}. There is a clear benefit in analytical tools that incorporate both spatio-temporal and particle-number degrees of freedom.

In the endeavor to combine particle-number degrees of freedom and spatio-temporal degrees of freedom, there are different approaches to choose from. One approach is to expand the states under investigation in terms of orthogonal bases. Operators can then be expressed in terms of overlaps among different bases elements. However, such an approach assumes the existence of a basis that is not only complete and orthogonal with respect to the particle-number degrees of freedom, but at the same time also complete and orthogonal with respect to the spatio-temporal degrees of freedom. To find such a basis is not a simple matter. It is the topic of this paper.

An alternative approach is to define states in terms of operators that would produce those states when they operate on the vacuum state. For normalized states, these would be unitary operators, often expressed as exponential operators, containing Hermitian operators in their exponents. The latter is often expressed as multivariate polynomials of creation and annihilation operators. It is particularly convenient if these polynomials are no higher than second order, thus naturally leading to the notion of Gaussian states \cite{weedbrook,contvar2}. Operations on states are represented by products of operators, which can be manipulated via the appropriate commutation relations. Due to the creation and annihilation operators, this approach is often associated with second quantization.

One way to incorporate spatio-temporal degrees of freedom in the latter approach, is to include multiple sets of creation and annihilation operators to represent different modes. It necessarily leads to a discrete set of modes representing the spatio-temporal degrees of freedom \cite{contvar1,contvar2}. However, it is not always desirable to perform analyses in terms of discrete modes. The result may require a truncation in the number of modes to allow explicit computations. Such a truncation can lead to large deviations between predictions and experimental results. For example, when discrete modes were used for the analysis of the evolution of biphoton states in turbulence \cite{ipe}, the resulting set of coupled differential equations had to be truncated before they could be solved. Moreover, the complexity of the analysis grew rapidly with the size of the truncated set. The predictions obtained from such a truncated set of equations gave a large disagreement when compared with experimental results \cite{oamturb}. It was only by performing the analysis in terms of the continuous plane-wave basis that this truncation problem could be overcome \cite{notrunc,qutrit}. Hence, it is preferable to incorporate spatio-temporal degrees of freedom into a particle-number formalism in such a way that it would allow one to use a continuous parameterization of both the particle-number degrees of freedom, as well as the spatio-temporal degrees of freedom.

Although we focus here mainly on the inclusion of the spatio-temporal degrees of freedom with the particle-number degrees of freedom, with a little more effort, we also incorporate the spin-degrees of freedom, thus exhausting all the degrees of freedom associated with the photon field. To do so, we make a slight simplification of the notation to avoid overly complex expressions.

In this paper, we'll derive the eigen-bases associated with so-called fixed-momentum quadrature operators. The resulting bases are called spatio-temporal quadrature bases. Although it makes heuristically sense that these bases would be complete and orthogonal, we proceed to show this explicitly. Since these bases incorporate the spatio-temporal degrees of freedom in terms of continuous degrees of freedom, they naturally lead to a functional formalism --- a path-integral approach \cite{peskin9,schulman,glimm}. However, by themselves, these spatio-temporal quadrature bases do not yet represent a fully-fledged formalism. They only provide the first step. The next step, which is to use these quadrature bases for the development of a generalized Wigner formalism, is beyond the scope of the current paper. However, we'll briefly discuss such a generalized Wigner formalism later.

The paper is organized as follows. In Sec.~\ref{prelim} we provide some background information and define convenient notation. The derivation of the spatio-temporal quadrature bases is provided in Sec.~\ref{fmqb}. Orthogonality and completeness conditions for these bases are considered in Secs.~\ref{mqbort} and \ref{fnalvol}, respectively. Finally, in Sec.~\ref{disc} a discussion and outlook are provided.

\section{\label{prelim}Preliminaries}

Before we address the main topic of this paper, namely the derivation of the spatio-temporal quadrature bases, we first consider some preliminary results that would be needed later. It also gives us an opportunity to define convenient notation and to show alternative approaches that do not work.

\subsection{\label{fmfs}Fixed-momentum Fock states}

The quantization of the electromagnetic field in the context of particle physics led to the definition of creation and annihilation operators $a_s^{\dag}({\bf k})$ and $a_s({\bf k})$ that carry spatio-temporal degrees of freedom in the form of the wave vector ${\bf k}$, which is proportional to the three-dimensional momentum vector, and spin degrees of freedom, represented by the spin index $s$. The creation and annihilation operators are assumed to obey a Lorentz covariant commutation relation
\begin{equation}
\left[ \hat{a}_s({\bf k}_1), \hat{a}_r^{\dag}({\bf k}_2) \right] = (2\pi)^3 \omega_1 \delta_{s,r} \delta({\bf k}_1-{\bf k}_2) ,
\label{commut}
\end{equation}
where $\delta_{s,r}$ is the Kronecker delta, $\delta({\bf k}_1-{\bf k}_2)$ is a (three-dimensional) Dirac delta function and the angular frequency $\omega_1$ is related to the wave vector via the dispersion relation $\omega_1=c|{\bf k}_1|$. The creation operators produce the elements of a momentum basis $a_s^{\dag}({\bf k})\ket{\rm vac}=\ket{{\bf k},s}$, which serves as a complete orthogonal basis for all single-photon states. It satisfies a Lorentz covariant orthogonality condition that reads
\begin{equation}
\braket{{\bf k}_1,r}{{\bf k}_2,s} = (2\pi)^3 \omega_1 \delta_{r,s} \delta({\bf k}_1-{\bf k}_2) .
\label{inprodk0}
\end{equation}

As a first attempt, one can generalize the single-photon momentum basis to a {\em fixed-momentum} Fock basis, where all photons in each basis element share the same spin and wave vector
\begin{align}
\begin{split}
\ket{n,{\bf k},s} & = \frac{1}{\sqrt{n!}} \left[a_s^{\dag}({\bf k})\right]^n \ket{{\rm vac}} \\
\bra{n,{\bf k},s} & = \frac{1}{\sqrt{n!}} \bra{{\rm vac}} \left[a_s({\bf k})\right]^n .
\end{split}
\label{fockfm}
\end{align}
where $n$ is the occupation number of photons in the state. The appropriate number operator for which these Fock states serve as eigenstates, with the occupation number $n$ being the eigenvalue, is given by
\begin{equation}
\hat{n} = \sum_r \int a_r^{\dag}({\bf k}') a_r({\bf k}')\ \frac{{\rm d}^3 k'}{(2\pi)^3\omega'} .
\label{numopk}
\end{equation}

At this point, we introduce a simpler notation to alleviate the complexity of subsequent expressions. To this end, integrals over momentum space with spin sums will henceforth be denoted by
\begin{equation}
\sum_s \int ...\ \frac{{\rm d}^3 k}{(2\pi)^3\omega} \rightarrow  \int ...\ \dbar_s k .
\end{equation}

Unfortunately, the attempt to define Fock states with fixed momenta leads to divergences. The inner product between two of these fixed-momentum Fock states with occupation numbers $m,n>1$, gives a product of Dirac delta functions with the same argument
\begin{align}
\braket{m,{\bf k}_1,r}{n,{\bf k}_2,s} & = \delta_{m,n} \left( \braket{{\bf k}_1,r}{{\bf k}_2,s} \right)^n \nonumber \\
& = \delta_{m,n} \left[ (2\pi)^3 \omega_1 \delta_{r,s} \delta({\bf k}_1-{\bf k}_2) \right]^n .
\end{align}
The product of Dirac delta functions will inevitably lead to unwanted divergences in any calculation. As a result, such an attempt to incorporate spatio-temporal degrees of freedom with the particle-number degrees of freedom fails. It follows that any other fixed-momentum basis (such as fixed-momentum quadrature bases or fixed-momentum coherent states) would suffer the same affliction.

\subsection{\label{fsfs}Fixed-spectrum Fock states}

A slight variation on the theme of fixed-momentum Fock states, is the notion of a fixed-{\em spectrum} Fock state, in which all the photons in the state share the same spectrum of plane waves. Making the presence of the spectrum explicit in the notation by the subscript $F$, we express these Fock states by
\begin{align}
\begin{split}
\ket{n_F} & = \frac{1}{\sqrt{n!}}\left({\hat{a}^{\dag}}_F\right)^n \ket{\rm vac} = \frac{\left(\ket{1_F}\right)^n}{\sqrt{n!}} \\
\bra{n_F} & = \frac{1}{\sqrt{n!}}\bra{\rm vac} \left(\hat{a}_F\right)^n = \frac{\left(\bra{1_F} \right)^n}{\sqrt{n!}} ,
\end{split}
\label{fsfsdef}
\end{align}
where the fixed-spectrum creation and annihilation operators are defined by
\begin{align}
\begin{split}
{\hat{a}^{\dag}}_F & = \int a_s^{\dag}({\bf k}) F_s({\bf k})\ \dbar_s k \\
\hat{a}_F & = \int F_s^*({\bf k}) a_s({\bf k})\ \dbar_s k .
\end{split}
\label{fsskepabs}
\end{align}
Fixed-spectrum single-photon state are expressed as
\begin{align}
\begin{split}
\ket{1_F} & = \int \ket{{\bf k},s} F_s({\bf k})\ \dbar_s k \\
\bra{1_F} & = \int F_s^*({\bf k}) \bra{{\bf k},s}\ \dbar_s k .
\end{split}
\end{align}
The fixed-spectrum Fock states also act as eigenstates of the number operator in Eq.~(\ref{numopk}): $\hat{n}\ket{n_F} = \ket{n_F} n$.

In all these expressions, the spectrum is represented by $F_s({\bf k})$, being a function of the three-dimensional wave vector ${\bf k}$ and the spin index $s$. It is normalized according to the expression
\begin{equation}
\int \left| F_s({\bf k}) \right|^2\ \dbar_s k = 1 .
\label{norm1}
\end{equation}
The normalized spectrum ensures that $\braket{1_F}{1_F}=1$ and $[a_F,{a^{\dag}}_F]=1$. As a result, the inner product between arbitrary Fock states with different spectra gives
\begin{equation}
\braket{m_F}{n_G} = \delta_{mn} (\langle F,G \rangle)^n .
\label{infockf}
\end{equation}
where
\begin{equation}
\langle F,G \rangle \equiv \int F_s^*({\bf k}) G_s({\bf k})\ \dbar_s k .
\label{binnespek}
\end{equation}
Hence, if we select a discrete set of spatial modes that is orthogonal and complete in terms of the spatio-temporal degrees of freedom to represent the spectra in these Fock states, we would obtain a set that obeys an orthogonality condition both in the spatio-temporal degrees of freedom, as well as in the particle-number degrees of freedom. Such a discrete fixed-spectrum Fock basis is denoted by $\{\ket{n_m}\}$, where $m$ is an index for the spatial-temporal modes.

Unfortunately, it would not be complete in both these degrees of freedom. Trying to expand a state consisting of the tensor product of two single-photon elements of the basis $\ket{\psi}=\ket{1_m}\ket{1_n}$, where $m\neq n$, one finds that all the inner products between $\ket{\psi}$ and the discrete fixed-spectrum Fock basis elements are zero. As a result, it cannot be represented in terms of the discrete Fock basis, which means that the discrete fixed-spectrum Fock basis is not complete. Again, our attempt fails.

\subsection{\label{padfock}Completeness of fixed-spectrum Fock basis}

For the moment, we'll ignore the orthogonality requirement and only focus on completeness. In what follows below, we'll eventually see that it makes sense to include all functions as spectra, and not only those that form an orthogonal basis for the spatial degrees of freedom. A completeness condition over such a space would naturally lead to a path-integral formulation \cite{peskin9,schulman,glimm}, because integrals over such a space are functional integrals that run over all the possible spectral functions in the space.

An expansion of an arbitrary pure state in terms of the full set of fixed-spectrum Fock states would have the form of a functional integral
\begin{equation}
\ket{\psi} = \sum_n \int \ket{n_F}{\cal C}_n[F]\ {\cal D}[F] ,
\end{equation}
where $F$ represents the complex functions for the spectra that define the fixed-spectrum Fock basis; the coefficient function ${\cal C}_n[F]$ is a functional (i.e., a function of functions); and the measure of the integral ${\cal D}[F]$ runs over all such functions $F$.

To investigate the completeness of the set of all fixed-spectrum Fock states, we'll consider the possibility to resolve the identity operator in terms of this set. For this purpose, we consider the functional integral for an operator given by
\begin{equation}
\hat{L} = \sum_n \int \ket{n_F} \bra{n_F}\ {\cal D}[F] .
\end{equation}
Using Eq.~(\ref{fsfsdef}), one can write it as
\begin{align}
\hat{L} & = \sum_n \int \frac{1}{n!}\left(\ket{1_F}\bra{1_F}\right)^n\ {\cal D}[F] \nonumber \\
& = \sum_n \int \frac{1}{n!}\left(\int\ket{{\bf k}_1,r} F_r({\bf k}_1)F_s^*({\bf k}_2) \bra{{\bf k}_2,s} \right. \nonumber \\
& \left. \times \dbar_r k_1\ \dbar_s k_2\right)^n\ {\cal D}[F] .
\label{Ldef}
\end{align}

To evaluate the functional integral, we interpret it as an ensemble averaging process, but with some differences. The majority of functions in the space over which the functional integral runs, would be similar to random functions. It is also reasonable to assume that the function values of most of these functions would be normally distributed. Those functions in the space that do not qualify as normally distributed random functions would form a subset of measure zero. To make this statement stronger than a mere assumption, one can specify it as a condition in the definition of the fixed-spectrum Fock basis. As a result, the functional integral of the product of a function with its complex conjugate, evaluated at different arguments would produce the equivalent of a completeness condition over the space of functions
\begin{equation}
\int F_r({\bf k}_1)F_s^*({\bf k}_2)\ {\cal D}[F] = (2\pi)^3 \omega_1 \delta_{r,s} \delta({\bf k}_1-{\bf k}_2) .
\label{condF}
\end{equation}
Note that there are two differences between what we have here and what is usually implied in delta-correlated random functions $\langle\chi(x_1)\chi^*(x_2)\rangle = \Delta \delta(x_1-x_2)$. The first difference is that the ensemble average always imply that the sum is divided by the number of elements. In the functional integral, there is no such division process implied. The second difference is that, to produce the singularity at the origin of the Dirac delta function, the random functions must have amplitudes of infinite magnitude. In the space of spectra that we consider, all functions are of finite energy, implying finite amplitudes. These two differences cancel each other to produce effectively the same result. Since we sum over an infinite number of finite amplitudes, but do not divide by the number of functions, we end up with a divergent value at the origin. It is also assumed that
\begin{equation}
\int F_r({\bf k}_1)F_s({\bf k}_2)\ {\cal D}[F] = \int F_r^*({\bf k}_1)F_s^*({\bf k}_2)\ {\cal D}[F] = 0 .
\label{condFcc}
\end{equation}
The functional integrals over products of more than two functions, either give zero for uneven numbers, or break up into a sum of products of integrals over just two functions for even numbers of functions, analogous to the ensemble averages of products of normally distributed random functions. For example
\begin{align}
\int & F_r({\bf k}_1)F_s^*({\bf k}_2)F_u({\bf k}_3)F_v^*({\bf k}_4)\ {\cal D}[F] \nonumber \\
 & = (2\pi)^6 \omega_1 \omega_3 \left[ \delta_{r,s} \delta_{u,v} \delta({\bf k}_1-{\bf k}_2) \delta({\bf k}_3-{\bf k}_4) \right. \nonumber \\
 & \left. + \delta_{r,v} \delta_{u,s} \delta({\bf k}_1-{\bf k}_4) \delta({\bf k}_3-{\bf k}_2) \right] .
\label{condF2}
\end{align}
In general, we can express the completeness condition to all orders as follows
\begin{align}
\int & \prod_{m=1}^{M} F({\bf k}_m;r_m) \prod_{n=1}^{N} F^*({\bf q}_n;s_n)\ {\cal D}[F] \nonumber \\
 & = \delta_{M,N} \sum_{\rm pert} \prod_{n=1}^{N} (2\pi)^3\omega_n \delta[r_n,s_{{\cal P}n}]\delta({\bf k}_n-{\bf q}_{{\cal P}n}) ,
\label{condals}
\end{align}
where the summation runs over all permutations and the subscript ${\cal P}n$ represents a permutation of all $N$ indices in the product. The Kronecker delta function for the spin is represented as $\delta[r,s]\equiv\delta_{r,s}$. We see that each function $F$ needs to be matched to a complex conjugate $F^*$ and vice versa. If any of these functions remains unmatched, the result is zero.

The relationship in Eq.~(\ref{condals}) serves as a generalized completeness condition for the space of all functions. It is now considered as a defining condition for the space of functions that defines the fixed-spectrum Fock basis.

As a result, the expression in Eq.~(\ref{Ldef}) becomes
\begin{equation}
\hat{L} = \sum_n\left(\int \ket{{\bf k},s} \bra{{\bf k},s}\ \dbar_s k \right)^n = \sum_n {\cal I}^n \equiv \mathds{1} ,
\label{LnaI}
\end{equation}
where ${\cal I}^n$ is a projection operator for $n$-particle states (an identity operator within the subspace of $n$-particle state) and $\mathds{1}$ is the identity operator for the entire space, including states with arbitrary numbers of particles. The representation succeeds as a resolution of the identity operator, which shows that, provided that the condition in Eq.~(\ref{condals}) is satisfied, the fixed-spectrum Fock basis is a complete basis for both the particle-number degrees of freedom and the spatio-temporal degrees of freedom
\begin{equation}
\sum_n \int \ket{n_F} \bra{n_F}\ {\cal D}[F] = \mathds{1} .
\label{fsfsvol}
\end{equation}

As a result, we obtained a successful completeness condition at the cost of orthogonality of the basis. Based on Eqs.~(\ref{condFcc}) and (\ref{fsfsvol}), we also have
\begin{equation}
\int \ket{m_F} \bra{n_F}\ {\cal D}[F] = \delta_{mn} {\cal I}^n ,
\label{fsfsvolx}
\end{equation}
which we'll need later.

\subsection{\label{fscoh}Fixed-spectrum coherent states}

In our pursuit for a complete orthogonal basis in spatio-temporal and particle-number degrees of freedom, any attempt to consider coherent state may seem like a waste of time, because one already knows that coherent states are not mutually orthogonal. However, we'll consider the completeness properties of fixed-spectrum coherent states for the benefit of later use.

The fixed-spectrum coherent states can be defined as an expansion in terms of fixed-spectrum Fock states
\begin{equation}
\ket{\alpha_F} = \exp \left(-\frac{1}{2}|\alpha|^2\right) \sum_{n=0}^{\infty} \frac{\alpha^n}{\sqrt{n!}}\ \ket{n_F} ,
\label{fskohdef}
\end{equation}
where $\alpha$ is a complex constant. In terms of the fixed-spectrum displacement operator, they are given by
\begin{equation}
\ket{\alpha_F} = \hat{D}(\alpha_F) \ket{\rm vac} = \exp\left(\hat{a}_{\alpha}^{\dag}-\hat{a}_{\alpha}\right) \ket{\rm vac}  ,
\label{verplfs}
\end{equation}
where
\begin{align}
\begin{split}
\hat{a}_{\alpha}^{\dag} & = \int \alpha_s({\bf k}) \hat{a}_s^{\dag}({\bf k})\ \dbar_s k \\
\hat{a}_{\alpha} & = \int \alpha^*_s({\bf k}) \hat{a}_s({\bf k})\ \dbar_s k .
\end{split}
\label{redefpq}
\end{align}
The complex function $\alpha_s({\bf k})$ is composed of the product of the complex parameter $\alpha$ and the normalized complex function $F_s({\bf k})$. Note that, according to Eq.~(\ref{norm1}),
\begin{align}
|| \alpha_s({\bf k}) ||^2 & \equiv \int \left| \alpha_s({\bf k}) \right|^2\ \dbar_s k \nonumber \\
 & = |\alpha|^2 \int \left| F_s({\bf k}) \right|^2\ \dbar_s k = |\alpha|^2 .
\label{normeta}
\end{align}

The inner product between different fixed-spectrum coherent states (i.e., where the photons in the respective coherent states in general have completely different spectra), can be calculated with the aid of Eq.~(\ref{infockf}). It reads
\begin{align}
\braket{\alpha_F}{\beta_G} & = \exp \left(-\frac{1}{2}|\alpha|^2-\frac{1}{2}|\beta|^2+\langle\alpha,\beta\rangle \right) \label{fsinprodf} \\
& = \exp \left(-\frac{1}{2}||\alpha_s({\bf k})-\beta_s({\bf k})||^2 \right) \exp(\im \Theta) , \nonumber
\end{align}
where, from Eq.~(\ref{normeta}),
\begin{equation}
||\alpha_s({\bf k})-\beta_s({\bf k})||^2 = \int \left|\alpha_s({\bf k})-\beta_s({\bf k})\right|^2\ \dbar_s k ,
\label{metric}
\end{equation}
gives the metric distance between the two spectra, and
\begin{equation}
\Theta = \Im\left\{ \langle\alpha,\beta\rangle \right\} = \Im\left\{\int \alpha_s^*({\bf k})\beta_s({\bf k})\ \dbar_s k \right\} ,
\label{kohfase}
\end{equation}
with $\Im\{\cdot\}$ giving the imaginary part of the argument.

Therefore, the inner product between different fixed-spectrum coherent states is related to the metric distance between their associated complex spectra in the space of functions. So, even when $\langle\alpha,\beta\rangle=0$, we still have $\braket{\alpha_F}{\beta_G}\neq 0$. Not surprisingly, it is not possible to define a basis in terms of such coherent states that are orthogonal. The space of functions for all fixed-spectrum coherent states forms a metric space (or normed vector space). The metric in the space of functions is related to the inner product between the associated fixed-spectrum coherent states, as
\begin{equation}
d\{\alpha,\beta\} \equiv ||\alpha_s({\bf k})-\beta_s({\bf k})||^2 = -\ln(|\braket{\alpha_F}{\beta_G}|^2) .
\end{equation}

To investigate the completeness of the fixed-spectrum coherent bases, we first consider
\begin{equation}
\hat{L} = \int \ket{\alpha_F}\bra{\beta_F}\ {\cal D}[F] ,
\end{equation}
where the complex constants $\alpha$ and $\beta$ are allowed to be different and where the functional integral only runs over the normalized functions $F$. Using the definition of the coherent states in Eq.~(\ref{fskohdef}) and applying Eq.~(\ref{fsfsvolx}), we get
\begin{equation}
\hat{L} = \exp \left(-\frac{1}{2}|\alpha|^2-\frac{1}{2}|\beta|^2+\alpha^*\beta {\cal I} \right) .
\end{equation}
By setting $\alpha=\beta$, we obtain
\begin{equation}
\int \ket{\alpha_F}\bra{\alpha_F}\ {\cal D}[F] = \exp \left(-|\alpha|^2+|\alpha|^2 {\cal I} \right) .
\end{equation}
Next, we also integrate over $\alpha$
\begin{align}
\int \ket{\alpha_F}\bra{\alpha_F}\ {\cal D}[\alpha_F] & = \int \exp \left(-|\alpha|^2+|\alpha|^2 {\cal I} \right)\ {\rm d}\alpha \nonumber \\
 & = \pi \sum_{n=0}^{\infty} {\cal I}^n = \pi \mathds{1} .
\end{align}
Hence, the functional integral over all complex functions $\alpha_F=\alpha F({\bf k})$ provide us with a completeness condition
\begin{equation}
\mathds{1} = \frac{1}{\pi} \int \ket{\alpha_F}\bra{\alpha_F}\ {\cal D}[\alpha_F] ,
\label{kohvol}
\end{equation}
which is analogous to what one obtains without the spatio-temporal degrees of freedom. The fixed-spectrum coherent states represent a (over-) complete basis, even though they are not orthogonal.

Since the identity is idempotent, we must have $\mathds{1}^2 = \mathds{1}$. Due to the non-orthogonality of the coherent states Eq.~(\ref{fsinprodf}), it leads to the awkward identity
\begin{align}
\mathds{1} & = \frac{1}{\pi^2} \int \ket{\alpha_F}\braket{\alpha_F}{\beta_G}\bra{\beta_G}\ {\cal D}[\alpha_F]\ {\cal D}[\beta_G] \nonumber \\
& = \frac{1}{\pi^2} \int \ket{\alpha_F} \exp\left(-\frac{1}{2}|\alpha|^2-\frac{1}{2}|\beta|^2+\langle\alpha,\beta\rangle \right) \bra{\beta_G} \nonumber \\
& \times {\cal D}[\alpha_F]\ {\cal D}[\beta_G] ,
\label{kohvol2}
\end{align}
which will come in handy later.

\subsection{\label{fsqb}Fixed-spectrum quadrature bases}

One can define fixed-spectrum quadrature operators directly in terms of the fixed-spectrum creation and annihilation operators
\begin{align}
\begin{split}
\hat{q}_F & = \frac{1}{\sqrt{2}} (\hat{a}_F+\hat{a}_F^{\dag}) \\
\hat{p}_F & = \frac{-\im}{\sqrt{2}} (\hat{a}_F-\hat{a}_F^{\dag}) .
\end{split}
\label{fsquadop}
\end{align}
Their commutation relation is given by
\begin{equation}
[\hat{q}_F,\hat{p}_G] = \im \Re\{\langle F,G \rangle\} ,
\label{fsquadkom}
\end{equation}
where $\Re\{\cdot\}$ represents the real part of the expression.

The fixed-spectrum quadrature basis elements are the eigenstates of the fixed-spectrum quadrature operators
\begin{align}
\begin{split}
\hat{q}_F\ket{q_F} & = \ket{q_F} q \\
\hat{p}_F\ket{p_F} & = \ket{p_F} p ,
\end{split}
\label{fseieqp}
\end{align}
and are assumed to be given by expansions
\begin{align}
\begin{split}
\ket{q_F} & = \sum_n \ket{n_F} \Theta_n(q_F) \\
\ket{p_F} & = \sum_n \ket{n_F} \Phi_n(p_F) ,
\end{split}
\label{fsquad}
\end{align}
in term of the fixed-spectrum Fock states. To find the expressions for the coefficient functions, we compute the overlaps
\begin{align}
\begin{split}
\Theta_n(q_F) & = \braket{n_F}{q_F} = \frac{1}{\sqrt{n!}} \bra{\rm vac} \left(\hat{a}_F\right)^n \ket{q_F} \\
\Phi_n(p_F) & = \braket{n_F}{p_F} = \frac{1}{\sqrt{n!}} \bra{\rm vac} \left(\hat{a}_F\right)^n \ket{p_F} ,
\end{split}
\label{koefreken}
\end{align}
where we used Eq.~(\ref{fsfsdef}) to express the fixed-spectrum Fock states in terms of fixed-spectrum annihilation operators. We now convert the products of annihilation operators into generating functions for such products
\begin{align}
\begin{split}
\sum_n \frac{\eta^n}{\sqrt{2^n n!}} \Theta_n(q_F) & = \bra{\rm vac} \exp\left(\frac{\eta}{\sqrt{2}}\hat{a}_F\right) \ket{q_F} \\
\sum_n \frac{(-\im)^n\eta^n}{\sqrt{2^n n!}} \Phi_n(p_F) & = \bra{\rm vac} \exp\left(\frac{-\im\eta}{\sqrt{2}}\hat{a}_F\right) \ket{p_F} ,
\end{split}
\label{koefreken0}
\end{align}
where $\eta$ is the generating parameter and where we introduced convenient constants in anticipation of our goal. Next, we exploit the fact that
\begin{equation}
\bra{\rm vac}\exp\left(K\hat{a}_F^{\dag}\right) = \bra{\rm vac}
\end{equation}
to insert exponentiated creation operators
\begin{align}
\begin{split}
\sum_n \frac{\eta^n}{\sqrt{2^n n!}} \Theta_n(q_F) & = \bra{\rm vac} \exp\left(\frac{\eta}{\sqrt{2}}\hat{a}_F^{\dag}\right) \\
& \times \exp\left(\frac{\eta}{\sqrt{2}}\hat{a}_F\right) \ket{q_F} \\
\sum_n \frac{(-\im)^n\eta^n}{\sqrt{2^n n!}} \Phi_n(p_F) & = \bra{\rm vac} \exp\left(\frac{\im\eta}{\sqrt{2}}\hat{a}_F^{\dag}\right) \\
& \times \exp\left(\frac{-\im\eta}{\sqrt{2}}\hat{a}_F\right) \ket{p_F} .
\end{split}
\label{koefreken1}
\end{align}
The exponential operators are combined using the Baker-Campbell-Hausdorff formula. The resulting combined exponential operators can be expressed in terms of fixed-spectrum quadrature operators. Hence,
\begin{align}
\begin{split}
\sum_n \frac{\eta^n}{\sqrt{2^n n!}} \Theta_n(q_F) & = \bra{\rm vac} \exp\left(\eta\hat{q}_F-\frac{\eta^2}{4}\right) \ket{q_F} \\
& = \Theta_0(q_F) \exp\left(\eta q -\frac{\eta^2}{4}\right) \\
\sum_n \frac{(-\im)^n\eta^n}{\sqrt{2^n n!}} \Phi_n(p_F) & = \bra{\rm vac} \exp\left(\eta\hat{p}_F-\frac{\eta^2}{4}\right) \ket{p_F} \\
& = \Phi_0(p_F) \exp\left(\eta p -\frac{\eta^2}{4}\right) ,
\end{split}
\label{koefreken2}
\end{align}
where we used the eigenvalue equations in Eq.~(\ref{fseieqp}) to pull the exponentiated quadrature operators through the basis elements. These results resemble the generating function for Hermite polynomials, given by
\begin{equation}
\exp(2x\nu-\nu^2) = \sum_{n=0}^{\infty} \frac{\nu^n}{n!} H_n(x) ,
\label{genherm}
\end{equation}
where $\nu$ is the generating parameter. The coefficient functions are therefore given by
\begin{align}
\begin{split}
\Theta_n(q_F) & = \frac{\Theta_0(q_F)}{\sqrt{2^n n!}} H_n(q) \\
\Phi_n(p_F) & = \frac{(\im)^n\Phi_0(p_F)}{\sqrt{2^n n!}} H_n(p) ,
\end{split}
\label{koefreken4}
\end{align}
in terms of Hermite polynomials, up to the zeroth order coefficient functions $\Theta_0(q)$ and $\Phi_0(p)$, which are yet to be determined. At this point, we note that the possible dependences on the normalized spectrum $F$ only appear in the zeroth order coefficient functions. So, if they only depend on the $q$- and $p$-parameters and not on the normalized spectrum, then the same would be true for all coefficient functions. Therefore, we assume that one can replace $q_F\rightarrow q$ and $p_F\rightarrow p$ in these expressions.

To find the expressions for the zeroth order coefficient functions, we consider the inner product between coefficient functions of different orders. Comparing the resulting expressions with the orthogonality condition for Hermite polynomials
\begin{equation}
\int H_m(x) H_n(x) \exp(-x^2)\ {\rm d}x = \sqrt{\pi} 2^n n!\ \delta_{mn} ,
\label{ortherm}
\end{equation}
we see that, if
\begin{align}
\begin{split}
|\Theta_0(q)|^2 & = \frac{1}{\sqrt{\pi}} \exp(-q^2) \\
|\Phi_0(p)|^2 & = 2\sqrt{\pi} \exp(-p^2)  ,
\end{split}
\end{align}
then the coefficient functions would obey the orthogonality conditions
\begin{align}
\begin{split}
\int \Theta_m(q) \Theta_n^*(q)\ {\rm d}q & = \delta_{mn} \\
\int \Phi_m(p) \Phi_n^*(p)\ {\rm d}p & = 2\pi\delta_{mn} .
\end{split}
\label{ortqpcf}
\end{align}
Assuming that the zeroth order coefficient functions are real-valued functions, we obtain expressions for the coefficient functions, given by
\begin{align}
\begin{split}
\Theta_n(q_F) & = \frac{1}{\pi^{1/4}\sqrt{2^n n!}} H_n(q) \exp\left(-\frac{q^2}{2}\right) \\
\Phi_n(p_F) & = \frac{(\im)^n\sqrt{2} \pi^{1/4}}{\sqrt{2^n n!}} H_n(p) \exp\left(-\frac{p^2}{2}\right) .
\end{split}
\label{koefpqn}
\end{align}
These coefficient functions are the same as those one has without the spatio-temporal degrees of freedom. In addition to the orthogonality conditions given in Eq.~(\ref{ortqpcf}), the coefficient functions of Eq.~(\ref{koefpqn}) also obey completeness conditions, given by
\begin{align}
\begin{split}
\sum_n \Theta_n^*(q) \Theta_n(q') & = \delta(q-q') \\
\sum_n \Phi_n^*(p) \Phi_n(p') & = 2\pi \delta(p-p') .
\end{split}
\label{volqpcf}
\end{align}

What is the effect of the spectra on the inner products among quadrature basis elements? Based on Eq.~(\ref{infockf}), the inner product between $q$-states with different spectra becomes
\begin{equation}
\braket{q_F}{q_G'} = \sum_n \mu^n \Theta_n^*(q) \Theta_n(q') ,
\end{equation}
where $\mu\equiv\langle F,G \rangle$. Using Eq.~(\ref{koefpqn}) and Mehler's formula \cite{mehler}, which is given by
\begin{align}
& \sum_{n=0}^{\infty} \frac{\rho^n}{2^n n!} H_n(x) H_n(y) \nonumber \\
& = \frac{1}{\sqrt{1-\rho^2}} \exp\left[ \frac{2\rho xy}{1-\rho^2} -\frac{(x^2+y^2)\rho^2}{1-\rho^2} \right] ,
\label{mehler}
\end{align}
we obtain an expression for the inner product that reads
\begin{align}
\braket{q_F}{q_G'} & = \frac{1}{\sqrt{\pi(1-\mu^2)}} \exp\left[ - \frac{\mu (q-q')^2}{1-\mu^2} \right] \nonumber \\
& \times \exp\left[ -\frac{(1-\mu)(q^2+{q'}^2)}{2(1+\mu)} \right] .
\label{mehqq}
\end{align}
For $G\rightarrow F$, but keeping the $q$'s different, we have $\mu\rightarrow 1$. We see that Eq.~(\ref{mehqq}) is singular for $\mu\rightarrow 1$. So we replace $\mu=1+\epsilon/2$ and consider the limit where $\epsilon\rightarrow 0$. It gives
\begin{equation}
\braket{q_F}{q_F'} = \lim_{\epsilon\rightarrow 0} \frac{1}{\sqrt{\pi\epsilon}} \exp\left[ -\frac{(q-q')^2}{\epsilon} \right] .
\end{equation}
One can show that the right-hand side represents a limit process for the Dirac delta function
\begin{equation}
\lim_{\epsilon\rightarrow 0} \frac{1}{\sqrt{\pi\epsilon}} \exp\left[ -\frac{(q-q')^2}{\epsilon} \right] = \delta(q-q') .
\end{equation}
Hence, we obtain the orthogonality condition
\begin{equation}
\braket{q_F}{q_F'} = \delta(q-q_0) .
\label{volgen}
\end{equation}
a similar condition applies for the $p$-basis.

On the other hand, if $F$ and $G$ are orthogonal, we have $\mu\rightarrow 0$, so that
\begin{equation}
\braket{q_F}{q_G'} \rightarrow \Theta_0^*(q) \Theta_0(q') = \frac{1}{\sqrt{\pi}} \exp\left[ \frac{-(q^2+{q'}^2)}{2} \right] .
\end{equation}
As a result, the fixed spectrum quadrature basis elements lose their orthogonality. Even if we use a discrete modal basis for the spatio-temporal degrees of freedom, we still do not obtain an orthogonal basis for both particle-number and spatio-temporal degrees of freedom. So, if we remain within the subspace associated with a specific spectral function, the fixed-spectrum quadrature bases are orthogonal bases, but beyond that they are not orthogonal.

\subsection{\label{padquad}Completeness of fixed-spectrum quadrature bases}

Considering the completeness of the fixed-spectrum quadrature bases, we first restrict the bases to the subspace of a specific spectral function and integrate over the $q$-parameter. We obtain
\begin{align}
\int \ket{q_F}\bra{q_F}\ {\rm d}q & = \sum_{m,n} \ket{m_F} \bra{n_F} \Theta_m(q) \Theta_n^*(q)\ {\rm d}q \nonumber \\
&  = \sum_n \ket{n_F} \bra{n_F} = {\cal I}_F ,
\end{align}
thanks to the orthogonality of the coefficient functions. The result represents a projection operator for states with arbitrary numbers of photons, but where their spatio-temporal degrees of freedom are defined by a specific spectral function $F$.

To investigate the completeness of the fixed-spectrum quadrature bases for the entire space of spectral functions, we need to employ functional integrals, as was done in Sec.~\ref{padfock}. First, we consider the case where the spectrum is the same, but the $q$-parameters are different
\begin{equation}
\hat{L} = \int \ket{q_F}\bra{q_F'}\ {\cal D}[F] .
\end{equation}
Substituting Eq.~(\ref{fsquad}), we obtain
\begin{align}
\hat{L} & = \sum_{mn} \int \ket{n_F} \Theta_m(q) \Theta_n^*(q') \bra{m_F}\ {\cal D}[F] \nonumber \\
& = \sum_n \Theta_n(q) \Theta_n^*(q')\ {\cal I}^n ,
\end{align}
where we used Eq.~(\ref{fsfsvolx}). It cannot be simplified further. However, if we also integrate over $q$ and use the orthogonality of the coefficient functions Eq.~(\ref{ortqpcf}), we get
\begin{align}
\int \hat{L}\ {\rm d}q & = \sum_n \int \Theta_n(q) \Theta_n^*(q')\ {\rm d}q\ {\cal I}^n \nonumber \\
& = \sum_n {\cal I}^n = \mathds{1} .
\end{align}
So, the entire space, including all normalized functions and all $q$-parameters, provides a complete basis. The same applies for the $p$-basis.

We are almost there. The fixed-spectrum quadrature bases are complete, but not orthogonal, unless we restrict them to subspaces for specific spectral functions.


\section{\label{fmqb}Eigenstates of fixed-momentum quadrature operators}

The stage is set to derive bases that are both complete and orthogonal with respect to the entire space of quantum states that can be defined in terms of both spatio-temporal degrees of freedom and particle-number degrees of freedom. These bases are what we'll call {\em spatio-temporal} quadrature bases.

For the derivation of the spatio-temporal quadrature bases, we start with the notion of {\em fixed-momentum} quadrature operators. These operators are directly defined in terms of the creation and annihilation operators for photonic states, which obey the Lorentz covariant commutation relation given in Eq.~(\ref{commut}). The fixed-momentum quadrature operators are given by
\begin{align}
\begin{split}
\hat{q}_s({\bf k}) & = \frac{1}{\sqrt{2}} \left[\hat{a}_s({\bf k})+\hat{a}_s^{\dag}({\bf k})\right] \\
\hat{p}_s({\bf k}) & = \frac{-\im}{\sqrt{2}} \left[\hat{a}_s({\bf k})-\hat{a}_s^{\dag}({\bf k})\right] .
\end{split}
\end{align}
They obey a Lorentz covariant commutation relation that reads
\begin{equation}
\left[ \hat{q}_r({\bf k}_1), \hat{p}_s({\bf k}_2) \right] = \im (2\pi)^3 \omega_1 \delta_{r,s} \delta({\bf k}_1-{\bf k}_2) .
\label{compqfm}
\end{equation}
The term {\em fixed-momentum} quadrature operator follows from the fact that they explicitly depend on the value of the wave vector.

It is now assumed that the fixed-momentum quadrature operators give rise to eigenstates and eigenvalue functions according to the following eigenvalue equations
\begin{align}
\begin{split}
\hat{q}_s({\bf k})\ket{q} & = \ket{q} q_s({\bf k}) \\
\hat{p}_s({\bf k})\ket{p} & = \ket{p} p_s({\bf k}) .
\end{split}
\label{eieqpfm}
\end{align}
The eigenvalue functions are real-valued functions because the quadrature operators are Hermitian. They can be separated into a real-valued magnitudes times real-valued normalized functions $q_s({\bf k})=q F_s({\bf k})$ and $p_s({\bf k})=p G_s({\bf k})$. For the sake of notational simplicity, we'll generally distinguish between the whole eigenvalue function and the magnitude simply by the presence or absence of the argument. The functions $F_s({\bf k})$ and $G_s({\bf k})$ are normalized in the sense
\begin{equation}
\int F_s^2({\bf k})\ \dbar_s k = \int G_s^2({\bf k})\ \dbar_s k = 1 .
\label{normfg}
\end{equation}

In the derivation that follows, we'll ignore the spin index to alleviate complexity of the expressions. After the derivation, we'll re-introduce the spin indices again.

It has been shown in Sec.~\ref{fmfs} that, when the eigenstates depend explicitly on the momentum, they would not be well-defined. Therefore, we'll assume that all the momentum dependences are integrated out. As a result, $\ket{q}\neq\ket{q({\bf k})}$ and $\ket{p}\neq\ket{p({\bf k})}$. Focusing on the first equation in Eq.~(\ref{eieqpfm}), we therefore start with an ansatz for $\ket{q}$ of the form
\begin{align}
\ket{q} & = \ket{{\rm vac}} V_0 + \int \ket{{\bf k}} V_1({\bf k}) \dbar k \nonumber \\
& + \frac{1}{2!} \int \ket{{\bf k}_a} \ket{{\bf k}_b} V_2({\bf k}_a,{\bf k}_b) \dbar k_a \dbar k_b \nonumber \\
& + \frac{1}{3!} \int \ket{{\bf k}_a} \ket{{\bf k}_b} \ket{{\bf k}_c} V_3({\bf k}_a,{\bf k}_b,{\bf k}_c) \dbar k_a \dbar k_b \dbar k_c + ...\ ,
\label{uitbq0}
\end{align}
where $V_n(\cdot)$ denotes coefficient functions to be determined. The subscript $n$ is not the spin; it represents the order of the term in which it appears. Applying the eigenvalue equation to the ansatz and solving the coefficient functions order by order, one obtains expressions for all coefficient functions with $n>0$ in terms of $V_0$. The first few are given by
\begin{align}
\begin{split}
V_1({\bf k}) & = \sqrt{2} V_0 q({\bf k}) \\
V_2({\bf k}_a,{\bf k}_b) & = 2 V_0 q({\bf k}_a) q({\bf k}_b) - V_0 \delta({\bf k}_a-{\bf k}_b) \\
V_3({\bf k}_a,{\bf k}_b,{\bf k}_c) & = 2\sqrt{2} V_0 q({\bf k}_a) q({\bf k}_b) q({\bf k}_c) \\
& - 3\sqrt{2} V_0 q({\bf k}_a) \delta({\bf k}_b-{\bf k}_c) ,
\end{split}
\label{koeff}
\end{align}
and so forth. Substituting these coefficient functions back into the ansatz in Eq.~(\ref{uitbq0}), we obtain an expression of the form
\begin{align}
\ket{q} & = \ket{{\rm vac}} V_0 + \left(\ket{Q} - \ket{R} \right) V_0 + \frac{1}{2!} \left(\ket{Q} - \ket{R} \right)^2 V_0 \nonumber \\
& + \frac{1}{3!} \left(\ket{Q} - \ket{R} \right)^3 V_0 + ... \nonumber \\
& = V_0 \sum_{m=0}^{\infty} \frac{1}{m!} \left(\ket{Q} - \ket{R} \right)^m \nonumber \\
& = V_0 \exp\left(\ket{Q} - \ket{R} \right) ,
\label{qkdef}
\end{align}
where $V_0$ is a global constant and
\begin{align}
\begin{split}
\ket{Q} & \equiv \sqrt{2} \int \ket{{\bf k},s} q_s({\bf k})\ \dbar_s k \\
\ket{R} & \equiv \frac{1}{2}\int \ket{{\bf k},s} \ket{{\bf k},s} \dbar_s k ,
\end{split}
\label{defqr}
\end{align}
with the spin indices being re-introduced. Note that
\begin{align}
\begin{split}
\hat{a}_s({\bf k}')\ket{Q} & = \ket{{\rm vac}} \sqrt{2} q_s({\bf k}') \\
\hat{a}_s({\bf k}')\ket{R} & = \ket{{\bf k}',s} ,
\end{split}
\end{align}
which implies that
\begin{align}
\begin{split}
\hat{a}_s({\bf k}')\exp(\ket{Q}) & = \exp(\ket{Q}) \sqrt{2} q_s({\bf k}') \\
\hat{a}_s({\bf k}')\exp\left(-\ket{R}\right) & = - \exp\left(-\ket{R}\right) \ket{{\bf k}',s} .
\end{split}
\label{anniQR}
\end{align}
The creation operator just adds a factor of $\ket{{\bf k}'}$. As an aside, it is interesting that $\exp\left(\ket{R}\right)$ acts like an `anti-vacuum' state in the sense that the application of an annihilation operator creates a ket-vector in the same way that it would normally be created by a creation operator acting on the vacuum state.

One can use the results in Eq.~(\ref{anniQR}) to test whether our solution satisfies the first eigenvalue equation in Eq.~(\ref{eieqpfm}):
\begin{align}
\hat{q}_s({\bf k})\ket{q} & = \frac{V_0}{\sqrt{2}} \left[ \exp(\ket{Q}) \exp\left(-\ket{R}\right) \sqrt{2} q_s({\bf k}) \right. \nonumber \\
& - \exp(\ket{Q}) \exp\left(-\ket{R}\right) \ket{{\bf k},s} \nonumber \\
& \left. + \exp(\ket{Q}) \exp\left(-\ket{R}\right) \ket{{\bf k},s}\right] \nonumber \\
& = \ket{q} q_s({\bf k}) .
\label{toetsqq}
\end{align}
It confirms that the expression for the eigenstates, given in Eq.~(\ref{qkdef}), satisfies the eigenvalue equation. However, the constant $V_0$ is still unspecified.

A similar procedure can be followed to obtain an expression for the eigenstate $\ket{p}$. It is given by
\begin{equation}
\ket{p} = W_0 \exp\left(\im\ket{P}+\ket{R}\right) ,
\label{pkdef}
\end{equation}
where $W_0$ is a global constant, $\ket{R}$ is given in Eq.~(\ref{defqr}) and
\begin{equation}
\ket{P} \equiv \sqrt{2} \int \ket{{\bf k},s} p_s({\bf k})\ \dbar_s k .
\end{equation}

\section{\label{mqbort}Orthogonality}

Since the fixed-momentum quadrature operators are Hermitian, it makes sense that the spatio-temporal quadrature bases would be orthogonal bases. However, what does it mean in the context of an infinite functional space? To investigate the orthogonality of these bases and to find expressions for $V_0$ and $W_0$, we need to compute $\braket{q}{q'}$ and $\braket{p}{p'}$. At the same time, it is helpful to compute $\braket{q}{p}$. For this purpose, we'll follow an operator approach, defining annihilation operators for $\ket{Q}$, $\ket{P}$ and $\ket{R}$, given by
\begin{align}
\begin{split}
\hat{a}_{Q} & = \sqrt{2} \int \hat{a}_s({\bf k}) q_s({\bf k})\ \dbar_s k \\
\hat{a}_{P} & = \sqrt{2} \int \hat{a}_s({\bf k}) p_s({\bf k})\ \dbar_s k \\
\hat{a}_{R} & = \frac{1}{2} \int \hat{a}_s({\bf k}) \hat{a}_s({\bf k})\ \dbar_s k .
\end{split}
\label{defqpra}
\end{align}
The corresponding creation operators are obtained as the adjoint operators, so that $\ket{Q}=\hat{a}_{Q}^{\dag} \ket{{\rm vac}}$, $\ket{P}=\hat{a}_{P}^{\dag} \ket{{\rm vac}}$ and $\ket{R}=\hat{a}_{R}^{\dag} \ket{{\rm vac}}$. Using these operators, we then define creation operators for the eigenstates
\begin{align}
\begin{split}
\hat{a}_q^{\dag} & = V_0 \exp\left(\hat{a}_{Q}^{\dag}-\hat{a}_{R}^{\dag}\right) \\
\hat{a}_p^{\dag} & = W_0 \exp\left(\im\hat{a}_{P}^{\dag}+\hat{a}_{R}^{\dag}\right) ,
\end{split}
\label{defqpskep}
\end{align}
so that $\ket{q}=\hat{a}_q^{\dag} \ket{{\rm vac}}$ and $\ket{p}=\hat{a}_p^{\dag} \ket{{\rm vac}}$. The overlaps among the eigenstates are then given by
\begin{align}
\begin{split}
\braket{q}{q'} & = \bra{{\rm vac}} \hat{a}_q \hat{a}_{q'}^{\dag} \ket{{\rm vac}} \\
\braket{p}{p'} & = \bra{{\rm vac}} \hat{a}_p \hat{a}_{p'}^{\dag} \ket{{\rm vac}} \\
\braket{q}{p} & = \bra{{\rm vac}} \hat{a}_q \hat{a}_p^{\dag} \ket{{\rm vac}} .
\end{split}
\label{qpoorv}
\end{align}

Starting with the last expression in Eq.~(\ref{qpoorv}), we obtain
\begin{align}
\braket{q}{p} & = V_0 W_0 \bra{{\rm vac}} \exp\left(\hat{a}_{Q}\right) \exp\left(-\hat{a}_{R}\right)\nonumber \\
& \times \exp\left(\im\hat{a}_{P}^{\dag}\right) \exp\left(\hat{a}_{R}^{\dag}\right) \ket{{\rm vac}} .
\label{fmqort0}
\end{align}
To evaluate this expression, one can rearrange the exponential operators in normal order. However, in the process new operators are generated in addition to the current ones. As a result, we expect to get an expression of the form
\begin{align}
& \exp\left(k_1\hat{a}_{Q}\right) \exp\left(k_2\hat{a}_{R}\right) \exp\left(k_3\hat{a}_{P}^{\dag}\right) \exp\left(k_4\hat{a}_{R}^{\dag}\right) \nonumber \\
& = \exp(h_0) \exp\left(h_1\hat{a}_{Q}^{\dag}\right) \exp\left(h_2\hat{a}_{P}^{\dag}\right) \exp\left(h_3\hat{a}_{R}^{\dag}\right) \nonumber \\
& \times \exp\left(h_4\hat{s}\right)\exp\left(h_5\hat{a}_{Q}\right)\exp\left(h_6\hat{a}_{P}\right) \exp\left(h_7\hat{a}_{R}\right)  ,
\label{omkeer0}
\end{align}
where $k_1$, $k_2$, $k_3$ and $k_4$ are assumed to be known constants; $h_0$, $h_1$, $h_2$, $h_3$, $h_4$, $h_5$, $h_6$ and $h_7$ are unknown constants; and
\begin{equation}
\hat{s} \equiv \frac{1}{2} \int \left[ \hat{a}_s^{\dag}({\bf k}) \hat{a}_s({\bf k}) + \hat{a}_s({\bf k}) \hat{a}_s^{\dag}({\bf k}) \right]\ \dbar_s k .
\label{numr}
\end{equation}
To obtain the required relationship, we follow a standard procedure where one introduces an auxiliary variable $t$ into the exponents on the left-hand side and converts the unknown constants into unknown functions of $t$ on the right-hand side
\begin{align}
& \exp\left(t k_1\hat{a}_{Q}\right) \exp\left(t k_2\hat{a}_{R}\right) \exp\left(t k_3\hat{a}_{P}^{\dag}\right) \exp\left(t k_4\hat{a}_{R}^{\dag}\right) \nonumber \\
& =  \exp[h_0(t)] \exp\left[h_1(t)\hat{a}_{Q}^{\dag}\right] \exp\left[h_2(t)\hat{a}_{P}^{\dag}\right] \exp\left[h_3(t)\hat{a}_{R}^{\dag}\right] \nonumber \\
& \times \exp\left[h_4(t)\hat{s}\right] \exp\left[h_5(t)\hat{a}_{Q}\right] \exp\left[h_6(t)\hat{a}_{P}\right] \exp\left[h_7(t)\hat{a}_{R}\right] .
\label{omkeert}
\end{align}
For consistency, the unknown functions must go to zero for $t=0$.

Next, we apply a derivative with respect to $t$ on both sides and then remove as many of the exponential operators as possible by operating with the respective inverse operators on the right-hand sides of both sides of the equation. The rather complicated expression that is obtained can be simplified with the aid of the identity
\begin{align}
\exp(\hat{X})\hat{Y}\exp(-\hat{X}) & = \hat{Y} + \left[\hat{X},\hat{Y}\right] + \frac{1}{2!}\left[\hat{X},\left[\hat{X},\hat{Y}\right]\right] \nonumber \\
& + \frac{1}{3!}\left[\hat{X},\left[\hat{X},\left[\hat{X},\hat{Y}\right]\right]\right] + ...\ ,
\label{eksopprod}
\end{align}
where $\hat{X}$ and $\hat{Y}$ are two arbitrary operators. In Appendix~\ref{fmquadcoms}, we provide all the necessary commutation relations to perform this task. The simplified expression can then be separated into different differential equations for all the unknown functions. Upon solving these differential equations, we obtain
\begin{align}
\begin{split}
h_0(t) & = \frac{\left( k_1^2 k_4 q^2 t + k_2 k_3^2 p^2 t + 2 k_1 k_3 \mu \right) t^2}{1-k_2 k_4 t^2} \\
h_1(t) & = \frac{k_1 k_4 t^2}{1-k_2 k_4 t^2} \\
h_2(t) & = \frac{k_3 t}{1-k_2 k_4 t^2} \\
h_3(t) & = \frac{k_4 t}{1-k_2 k_4 t^2} \\
h_4(t) & = -\ln\left(1-k_2 k_4 t^2\right) \\
h_5(t) & = \frac{k_1 t}{1-k_2 k_4 t^2} \\
h_6(t) & = \frac{k_2 k_3 t^2}{1-k_2 k_4 t^2} \\
h_7(t) & = \frac{k_2 t}{1-k_2 k_4 t^2} .
\end{split}
\label{oplh}
\end{align}
where $\mu=\langle q,p\rangle$ is the inner product between the eigenvalue functions. One can now substitute the functions back into Eq.~(\ref{omkeert}). For the case under consideration, as given in Eq.~(\ref{fmqort0}), we substitute $k_1=1$, $k_2=-1$, $k_3=\im$ and $k_4=1$. Then we set $t=1$ to obtain the expression for the product of exponential operators in normal order
\begin{align}
\hat{a}_q \hat{a}_p^{\dag} & = V_0 W_0 \exp\left(\frac{q^2}{2} + \frac{p^2}{2} + \im \mu \right) \exp\left(\frac{\hat{a}_{Q}^{\dag}}{2}\right) \nonumber \\
& \times \exp\left(\frac{\im\hat{a}_{P}^{\dag}}{2}\right) \exp\left(\frac{\hat{a}_{R}^{\dag}}{2}\right) \exp\left[-\ln(2)\hat{s}\right] \nonumber \\
& \times \exp\left(\frac{\hat{a}_{Q}}{2}\right) \exp\left(\frac{-\im\hat{a}_{P}}{2}\right) \exp\left(-\frac{\hat{a}_{R}}{2}\right) .
\label{omkeerfin}
\end{align}
The expression for the overlap follows by contracting the vacuum state on both sides. For this purpose, we note that $\exp(K\hat{a}_{Q})\ket{\rm vac} = \ket{\rm vac}$, regardless of the value of $K$. The same applies for $\hat{a}_{P}$ and $\hat{a}_{R}$. Using the normal ordered form for $\hat{s}$, as given in Eq.~(\ref{nonumr}), we obtain
\begin{equation}
\bra{\rm vac}\exp[-\ln(2)\hat{s}]\ket{\rm vac} = \exp[-\ln(2)\Omega] = \frac{1}{2^{\Omega}} ,
\end{equation}
where $\Omega$ is defined in Eq.~(\ref{nonumr}). So, the overlap becomes
\begin{equation}
\braket{q}{p} = \frac{V_0 W_0}{2^{\Omega}} \exp\left(\frac{q^2}{2} + \frac{p^2}{2} + \im \mu \right) .
\label{fmqort1}
\end{equation}
We now define the global constants as
\begin{align}
\begin{split}
V_0 & = 2^{\Omega/2} \exp\left(-\frac{q^2}{2}\right) \\
W_0 & = 2^{\Omega/2} \exp\left(-\frac{p^2}{2}\right) .
\end{split}
\label{fg0def}
\end{align}
As a result, the overlap becomes
\begin{equation}
\braket{q}{p} = \exp\left( \im \mu \right)  = \exp\left[ \im \int q_s({\bf k}) p_s({\bf k}) \dbar_s k \right] .
\label{oorvqp}
\end{equation}

To consider the other two overlaps in Eq.~(\ref{fmqort0}), we notice that they can be obtained from the same result with the appropriate substitutions. For $\braket{q}{q'}$, we need to replace $P\rightarrow Q'$ and $p\rightarrow q'$. The inner product now represents $\mu \rightarrow \langle q,q' \rangle$. Using these replacements, together with the appropriate assignment for the constants: $k_1=1$, $k_2=-1$, $k_3=1$ and $k_4=-1$, we obtain
\begin{align}
\begin{split}
h_0(t) & = \frac{- \left[ q^2 t + (q')^2 t - 2 \mu \right] t^2}{1-t^2} \\
h_1(t) & = h_6(t) = \frac{- t^2}{1-t^2} \\
h_2(t) & = -h_3(t) = h_5(t) = -h_7(t) = \frac{ t}{1-t^2} \\
h_4(t) & = -\ln\left(1-t^2\right)  .
\end{split}
\label{oplqq}
\end{align}
These functions are all singular at $t=1$. Therefore, we need to consider the overlap as a limit. For this purpose, we substitute $t=1-\epsilon$ and Eq.~(\ref{fg0def}). The resulting overlap then becomes
\begin{equation}
\braket{q}{q'} = \lim_{\epsilon\rightarrow 0} \frac{1}{\epsilon^{\Omega}} \exp\left\{ \frac{- \left[ q^2 + (q')^2 - 2 \mu \right]}{2\epsilon} \right\} .
\label{oorvqq}
\end{equation}
Note that
\begin{equation}
q^2 + (q')^2 - 2 \mu = ||q_s({\bf k})-q_s'({\bf k})||^2 \geq 0 ,
\end{equation}
where
\begin{equation}
||f_s({\bf k})||^2 \equiv \int f_s^2({\bf k})\ \dbar_s k .
\label{normdef}
\end{equation}
For $||q_s({\bf k})-q_s'({\bf k})||^2 > 0$, the limit in Eq.~(\ref{oorvqq}) gives zero and for $||q_s({\bf k})-q_s'({\bf k})||^2 = 0$ the limit gives infinity. Hence, the result behaves as a Dirac delta {\em functional}, which runs over an infinity number of degrees of freedom. It enforces the equality of the eigenvalue functions $q({\bf k})$ and $q'({\bf k})$. We can thus express the overlap as
\begin{equation}
\braket{q}{q'} = \delta[q_s({\bf k})-q_s'({\bf k})] ,
\label{oorvqq0}
\end{equation}
where we ignored some unspecified constant factor. A similar expression applies for the remaining overlap
\begin{equation}
\braket{p}{p'} = \delta[p_s({\bf k})-p_s'({\bf k})] .
\label{oorvpp0}
\end{equation}
The expressions in Eqs.~(\ref{oorvqq0}) and (\ref{oorvpp0}) represent the required orthogonality conditions for the spatio-temporal quadrature bases.

\section{\label{fnalvol}Completeness}

To investigate the completeness of the spatio-temporal quadrature bases, we consider an operator defined by
\begin{equation}
\hat{B} = \int \ket{q} \bra{q}\ {\cal D}[q] .
\end{equation}
Having shown in Eq.~(\ref{kohvol}) that the fixed-spectrum coherent states can be used to resolve the identity, we operate on both side with identities resolved in terms of coherent states
\begin{align}
\hat{B} & = \int \mathds{1} \ket{q} \bra{q} \mathds{1}\ {\cal D}[q] \nonumber \\
& = \frac{1}{\pi^2} \int \ket{\alpha_F} \braket{\alpha_F}{q} \braket{q}{\beta_G} \bra{\beta_G}\ {\cal D}[q]\ {\cal D}[\alpha_F]\ {\cal D}[\beta_G] .
\end{align}
To develop the expression further, we need the overlap between the quadrature basis and the fixed-spectrum coherent states. It is computed in Appendix~\ref{fopi}. After substituting the expressions for these overlaps Eq.~(\ref{oorvqkoh0}) into the above expression, we evaluate the functional integration over $q$. The result is
\begin{align}
\hat{B} & = \frac{\kappa}{\pi^2} \int \ket{\alpha_F} \exp\left\{- \int \frac{1}{2} \left|\alpha_s({\bf k})\right|^2- \alpha_s^*({\bf k})\beta_s({\bf k}) \right. \nonumber \\
& \left. + \frac{1}{2} \left|\beta_s({\bf k})\right|^2\ {\rm d}k \right\} \bra{\beta_G}\ {\cal D}[\alpha_F]\ {\cal D}[\beta_G] ,
\end{align}
where $\kappa$ as an unspecified constant. Comparing this result with Eq.~(\ref{kohvol2}), we see that it is proportional to an identity operator. Hence,
\begin{equation}
\frac{1}{\kappa} \int \ket{q} \bra{q}\ {\cal D}[q] = \mathds{1} .
\label{fquadqvol}
\end{equation}
A similar procedure gives
\begin{equation}
\frac{1}{\kappa} \int \ket{p} \bra{p}\ {\cal D}[p] = \mathds{1} .
\label{fquadpvol}
\end{equation}
The expressions in Eqs.~(\ref{fquadqvol}) and (\ref{fquadpvol}) imply that the spatio-temporal quadrature bases do indeed obey completeness conditions.

\section{\label{disc}Discussion and outlook}

In our endeavor to find bases that incorporate both particle-number degrees of freedom and spatio-temporal degrees of freedom, we found that the eigenstates of fixed-momentum quadrature operators satisfy our requirements. These quadrature bases are complete and orthogonal with respect to both particle-number degrees of freedom and spatio-temporal degrees of freedom. We refer to these as spatio-temporal quadrature bases, because the term `quadrature' already gives reference to the particle-number degrees of freedom.

We derived expressions for the spatio-temporal quadrature basis elements, both in terms of integrals over the single-photon momentum basis and in terms of operators that create these basis elements from the vacuum. The latter allows us to compute the overlaps among elements of these bases to show that they give rise to the notion of Dirac delta functionals, which represent orthogonality conditions for the bases. Using functional integrals, we also showed that the spatio-temporal quadrature bases satisfy completeness conditions.

One may get the impression that the spatio-temporal quadrature bases are very similar to the fixed-spectrum quadrature bases. The only real difference is the appearance of $\ket{R}$ inside the expressions for the spatio-temporal quadrature bases. Yet, it is quite easy to show that the fixed-spectrum quadrature basis elements are not able to serve as eigenstates of the fixed-momentum quadrature operators, according to the eigenvalue equations in Eq.~(\ref{eieqpfm}). What is the role of $\ket{R}$ that could lead to such a significant difference between these bases? To demonstrate the role of $\ket{R}$, one can try to expand a spatio-temporal quadrature basis element in terms of a fixed-spectrum Fock basis for a spectral function given by the eigenvalue function $q({\bf k})$ for that element. Such an expansion only reproduces the equivalent fixed-spectrum quadrature basis element (without the $\ket{R}$) and not the spatio-temporal quadrature basis element (with the $\ket{R}$). Hence, we conclude that the fixed-spectrum quadrature basis elements are the projections of the spatio-temporal quadrature basis elements onto the subspaces associated with the spectral functions given by $q({\bf k})$. The presence of $\ket{R}$ in the expressions of the spatio-temporal quadrature bases imply that their elements extend beyond the subspaces defined by their spectral functions $q({\bf k})$.

With the spatio-temporal quadrature bases in hand, one can formulate powerful analytical tools to investigate the evolution of quantum states that incorporate all the degrees of freedom that photonic states can possess. The idea is to generalize Wigner functions, which represent the particle-number degrees of freedom, to the notion of Wigner functionals, which incorporate both particle-number degrees of freedom and spatio-temporal degrees of freedom. Formally, such a Wigner functional would be defined by a functional integral
\begin{align}
W[q,p] \equiv & \int \bra{q+\frac{x}{2}}\hat{\rho}\ket{q-\frac{x}{2}}\nonumber \\
& \times \exp\left[ -\im \int p({\bf k}) x({\bf k}) {\rm d}k \right]\ {\cal D}[x] ,
\end{align}
based on the spatio-temporal $q$-basis. Such Wigner functionals would then be able to represent not only quantum states that depend on all these degrees of freedom, but also operators that incorporate all these degrees of freedom. As such, the formalism would be useful for measurements of states with arbitrary numbers of photons that incorporate spatio-temporal degrees of freedom \cite{sqltreps,lvovsky0,boydsupres,chille,supres,qsense}. One would compute the predicted outcomes from such measurements with a functional integral
\begin{equation}
\langle \hat{A} \rangle = \int W_{\hat{A}}[q,p] W_{\hat{\rho}}[q,p]\ {\cal D}[q,p] ,
\end{equation}
where $\hat{A}$ represents the Hermitian operator for the measurement.

Apart from the Wigner functionals, one can also consider the generalization of other quasi-distributions, such as the Glauber-Sudarshan $P$-distribution \cite{glauber,sudarshan} or the Husimi $Q$-distribution \cite{husimi}. For such cases, one may employ the fixed-spectrum coherent states, leading to functional integral expressions. For instance, a density operator would be represented in terms of the functional $P$-distribution by
\begin{equation}
\hat{\rho} = \int \ket{\alpha_F} P[\alpha_F] \bra{\alpha_F}\ {\cal D}[\alpha_F] .
\end{equation}
These quasi-distributions would be related to the Wigner functionals via functional integral expressions.

Such a formalism based on Wigner functionals (or functional quasi-distributions) is current still a work in progress and is therefore beyond the scope of the current paper. The hope is eventually to develop a tool that would enable one to study physical situations in quantum optics where multiple degrees of freedom are playing significant roles in what is being observed.


\appendix

\section{\label{fmquadcoms}Commutation relations}

Here, we derive the commutation relations that are associated with the operators in Eq.~(\ref{defqpra}). Our starting point is the commutation relation for $\hat{a}({\bf k})$ and $\hat{a}^{\dag}({\bf k})$, given in Eq.~(\ref{commut}).

If $[\hat{X},\hat{Y}]=\hat{Z}$, then $[\hat{Y}^{\dag},\hat{X}^{\dag}]=\hat{Z}^{\dag}$. So, we do not provide commutation relations that can be obtained directly from others by performing an adjoint operation.

We group the commutation relations in batches. First, all annihilation operators commute among themselves
\begin{align}
\begin{split}
\left[\hat{a}_s({\bf k}),\hat{a}_{Q}\right] & = \left[\hat{a}_s({\bf k}),\hat{a}_{P}\right] = \left[\hat{a}_s({\bf k}),\hat{a}_{R}\right] = 0 \\
\left[\hat{a}_{Q},\hat{a}_{Q}\right] & = \left[\hat{a}_{P},\hat{a}_{P}\right] = \left[\hat{a}_{R},\hat{a}_{R}\right] = 0 \\
\left[\hat{a}_{Q},\hat{a}_{P}\right] & = \left[\hat{a}_{Q},\hat{a}_{R}\right] = \left[\hat{a}_{P},\hat{a}_{R}\right] = 0 .
\end{split}
\label{komqpr0}
\end{align}
Unless the relation contains both $\hat{a}_{Q}$ and $\hat{a}_{P}$, as in
\begin{equation}
\left[\hat{a}_{Q},\hat{a}_{P}^{\dag}\right] = 2 \int q_s({\bf k}) p_s({\bf k})\dbar_s k \equiv 2 \mu ,
\label{komqp}
\end{equation}
we do not henceforth, show the commutation relation for them both since they are the same. In Eq.~(\ref{komqp}), $\mu$ represents the inner product between the functions
\begin{equation}
\mu = \langle q,p\rangle \equiv \int q_s({\bf k})p_s({\bf k})\ \dbar_s k .
\label{binneprod}
\end{equation}

The next batch involves $\hat{a}^{\dag}({\bf k})$
\begin{align}
\begin{split}
\left[\hat{a}_{Q},\hat{a}_s^{\dag}({\bf k})\right] & = \sqrt{2} q_s({\bf k}) \\
\left[\hat{a}_{P},\hat{a}_s^{\dag}({\bf k})\right] & = \sqrt{2} p_s({\bf k}) \\
\left[\hat{a}_{R},\hat{a}_s^{\dag}({\bf k})\right] & = \hat{a}_s({\bf k}) .
\end{split}
\label{komqprk}
\end{align}

Then, we have relations that include the $R$-operators
\begin{align}
\begin{split}
\left[\hat{a}_{R},\hat{a}_{Q}^{\dag}\right] & = \hat{a}_{Q} \\
\left[\hat{a}_{R},\hat{a}_{R}^{\dag}\right] & = \hat{s} ,
\end{split}
\label{komqpr3}
\end{align}
where $\hat{s}$ is the symmetrized number operator, given in Eq.~(\ref{numr}). It can also be written in normal ordered form
\begin{equation}
\hat{s} = \int \hat{a}_s^{\dag}({\bf k}) \hat{a}_s({\bf k})\ \dbar_s k + \frac{1}{2} \int \delta(0)\ \dbar_s k = \hat{n} + \Omega ,
\label{nonumr}
\end{equation}
where $\Omega$ is a divergent constant. The commutation relations involving the symmetrized number operator are
\begin{align}
\begin{split}
\left[\hat{a}_s({\bf k}),\hat{s}\right] & = \hat{a}_s({\bf k}) \\
\left[\hat{a}_{Q},\hat{s}\right] & = \hat{a}_{Q} \\
\left[\hat{a}_{R},\hat{s}\right] & = 2\hat{a}_{R} .
\end{split}
\label{komnqpr}
\end{align}

\section{\label{fopi}Quadrature representations of fixed-spectrum coherent states}

Here, we consider how to expand fixed-spectrum coherent states in terms of the spatio-temporal quadrature bases. For this purpose, we employ the eigenstate property of the coherent states:
\begin{align}
\begin{split}
\hat{a}_{Q}\ket{\alpha_F} & = \ket{\alpha_F} \sqrt{2} \langle q,\alpha\rangle \\
\hat{a}_{R}\ket{\alpha_F} & = \ket{\alpha_F} \frac{1}{2} {\cal A}_2 ,
\end{split}
\end{align}
where
\begin{equation}
{\cal A}_2 \equiv \int \alpha_s^2({\bf k})\ \dbar_s k .
\end{equation}
Therefore,
\begin{align}
\braket{q}{\alpha_F} & = V_0 \bra{\rm vac} \exp(\hat{a}_{Q}-\hat{a}_{R}) \ket{\alpha_F} \nonumber \\
 & = V_0 \braket{\rm vac}{\alpha_F} \exp\left[ \sqrt{2} \langle q,\alpha\rangle-\frac{1}{2} {\cal A}_2\right] .
\label{oorvqkoh}
\end{align}
If we express $\alpha_s({\bf k})$ in its real and imaginary parts,
\begin{equation}
\alpha_s({\bf k}) \rightarrow \frac{1}{\sqrt{2}}\left[q_{s0}({\bf k})+\im p_{s0}({\bf k})\right] ,
\end{equation}
and use Eq.~(\ref{fg0def}), we obtain
\begin{align}
\braket{q}{\alpha_F} & = 2^{\Omega/2} \exp\left\{- \int \frac{1}{2} \left[q_s({\bf k})-q_{s0}({\bf k})\right]^2 \right. \nonumber \\
 & \left. - \im p_{s0}({\bf k}) \left[q_s({\bf k})-\frac{1}{2}q_{s0}({\bf k})\right]\ \dbar_s k \right\} .
\label{oorvqkoh0}
\end{align}


\end{document}